\documentclass[10pt,journal,doublecolumn,twoside]{IEEEtran}
\IEEEoverridecommandlockouts

\usepackage{subfigure}
\usepackage{colortbl}
\usepackage{bm}

\usepackage{graphicx}  
\usepackage{url}       

\usepackage{amsmath}   
\usepackage{cite}

\usepackage{amsfonts,amssymb}

\usepackage{stfloats}

\usepackage{cases}
\usepackage{algorithm}
\usepackage{multirow}
\usepackage{algorithmic}

\newcommand{\ls}[1]
    {\dimen0=\fontdimen6\the\font
     \lineskip=#1\dimen0
     \advance\lineskip.5\fontdimen5\the\font
     \advance\lineskip-\dimen0
     \lineskiplimit=.9\lineskip
     \baselineskip=\lineskip
     \advance\baselineskip\dimen0
     \normallineskip\lineskip
     \normallineskiplimit\lineskiplimit
     \normalbaselineskip\baselineskip
     \ignorespaces
    }

\newcounter{TempEqCnt}

\usepackage[left=0.68in,right=0.70in,top=0.75in,bottom=0.6in]{geometry}

\begin{document}
\title{Mode Hopping for Anti-Jamming in Cognitive Radio Networks}
\author{Wenchi Cheng$^{\dagger}$, Zan Li$^{\dagger}$, Feifei Gao$^{\ddagger}$, Liping Liang$^{\dagger}$, and Hailin Zhang$^{\dagger}$

\IEEEauthorblockA{$^{\dagger}$State Key Laboratory of Integrated
Services Networks, Xidian University, Xi'an, China\\
$^{\ddagger}$Tsinghua National Laboratory for Information Science and Technology (TNList)\\
E-mail: \{\emph{wccheng@xidian.edu.cn}, \emph{zanli@xidian.edu.cn}, \emph{feifeigao@ieee.org}\}}

\thanks{\ls{.5}
This work was supported in part by the National Natural Science Foundation of China (No. 61771368) and Young Elite Scientists Sponsorship Program by CAST (2016QNRC001).
}

}

\maketitle

\begin{abstract}
Frequency hopping (FH) is an effective anti-jamming technology in cognitive radio networks (CRNs). However, it is difficult to significantly increase the anti-jamming results because of the growing crowded spectrum in wireless communications. Orbital momentum angular (OAM) provides a new mode dimension for anti-jamming without consuming extra power and frequency resources in CRNs. In this paper, we propose the mode-frequency hopping (MFH) scheme, which jointly uses the mode hopping and the traditional FH scheme for anti-jamming to significantly increase the capacity of secondary users (SUs). 
We derive the false alarm probability and transmission outage probability for CRNs under our proposed MFH scheme. Then, based on the two derived probabilities, we calculate the capacity of SUs under the MFH scheme. Numerical results show that our developed MFH scheme can achieve better anti-jamming results as compared with the traditional FH scheme in CRNs.

\end{abstract}

\section{Introduction}
\IEEEPARstart{C}{ognitive} radio (CR) is a promising technology for efficient spectrum efficiency~\cite{CR_spect,2017_VANET} and with wide range of applications~\cite{CR_appli,2017_Big}. In cognitive radio networks (CRNs), when the channel is not occupied by the primary users (PUs), the secondary users (SUs) are allowed to access the licensed idle channel for their own transmissions. However, because of no licensed spectrum for SUs, highly dynamic spectrum activities of PUs, and distribution of SUs, CRNs face severe security challenges.

Academic researchers have paid much attention to the security issues in CRNs~\cite{2011_emulation,2012_emulation,2012_attack}. 
Generally, PU emulation, eavesdropping attack, and jamming attack are major attacks in CRNs. The PU emulation is that the SUs disguise the PUs and then access the licensed channels to transmit their own signals~\cite{2011_emulation,2012_emulation}. Thus, the  required throughput of PUs cannot be guaranteed in CRNs. A defense strategy using belief propagation has been proposed against the PU emulation attacker in CRNs, thus detecting effectively the PU emulation attackers~\cite{2012_emulation}. The jamming attacks, which make impact on the availability of networks, send signals to the licensed channels as interference without taking into account the existence of SUs or PUs, thus resulting in the useless of the channel for both PUs and SUs~\cite{2012_attack}. By modeling the interaction between SUs and attackers, a defense strategy using the Markov decision process has been studied~\cite{2012_attack}.

When the channel is jammed by attackers, we can use the channel switching based on defense strategies in medium access control (MAC) layer~\cite{2012_emulation,2012_attack} or the frequency hopping (FH) in physical layer for anti-jamming in CRNs~\cite{2014_FH,2010_FH_Ad,2009_FH,2015_FH}. A novel architecture has been proposed to implement an adaptive FH cognitive radio for current spectrum-limited environment~\cite{2014_FH}. In~\cite{2010_FH_Ad}, the proposed adaptive multiple rendezvous control channel based on FH aims at reducing the time to rendezvous and increasing the overall network performance. In addition, a few types of FH schemes and their properties have been provided~\cite{2015_FH}. However, FH requires a relatively wide bandwidth to hop. The growing crowded spectrum makes FH difficult to guarantee the reliability of CRNs in wireless networks. Hence, how to significantly increase the anti-jamming results in physical layer of CRNs remains a critical and open challenge.

Orbital angular momentum (OAM), which describes the helical phase front of electromagnetic waves, has the potential to achieve high spectrum efficiency~\cite{2013_spec,2018_spec} and efficient anti-jamming results without consuming extra power and frequency resources in wireless communications~\cite{2018_MFH}. Due to the orthogonality among different OAM-modes, the proposed mode hopping used for anti-jamming can achieve the same bit error rate (BER) within the narrow band in comparison with the traditional FH scheme. Also, the joint mode hopping and the traditional FH scheme, which is called mode-frequency hopping (MFH) scheme, can significantly decrease the BER of radio vortex wireless communications~\cite{2018_MFH}.

To further enhance the anti-jamming results using the new mode dimension for CRNs, in this paper we propose to use MFH scheme for CRNs to increase the capacity of SUs. First, based on the instantaneous signal-to-interference-plus-noise ratio (SINR) sensing threshold and transmission outage threshold, we derive the false alarm probability corresponding to the sensing phase and the transmission outage probability corresponding to the transmission phase, respectively. Then, based on the two derived probabilities, we calculate the capacity of SUs in CRNs under our proposed MFH scheme. We conduct extensive numerical results to evaluate our developed schemes, showing that the probability jammed by attacker decreases and our developed MFH scheme can achieve higher capacity for SUs than the traditional FH scheme.

The remainder of this paper is organized as follows. Section~\ref{sec:sys} gives the MFH scheme based system model for CRNs. Section~\ref{sec:anti} derives the false alarm probability, transmission outage probability, and the capacity of SUs in CRNs under our proposed MFH scheme. Section~\ref{sec:perfor} evaluates the MFH scheme and compares it with the traditional FH in CRNs. The paper concludes with Section~\ref{sec:conc}.

\section{System Model}\label{sec:sys}
\begin{figure}
\centering
\includegraphics[scale=0.75]{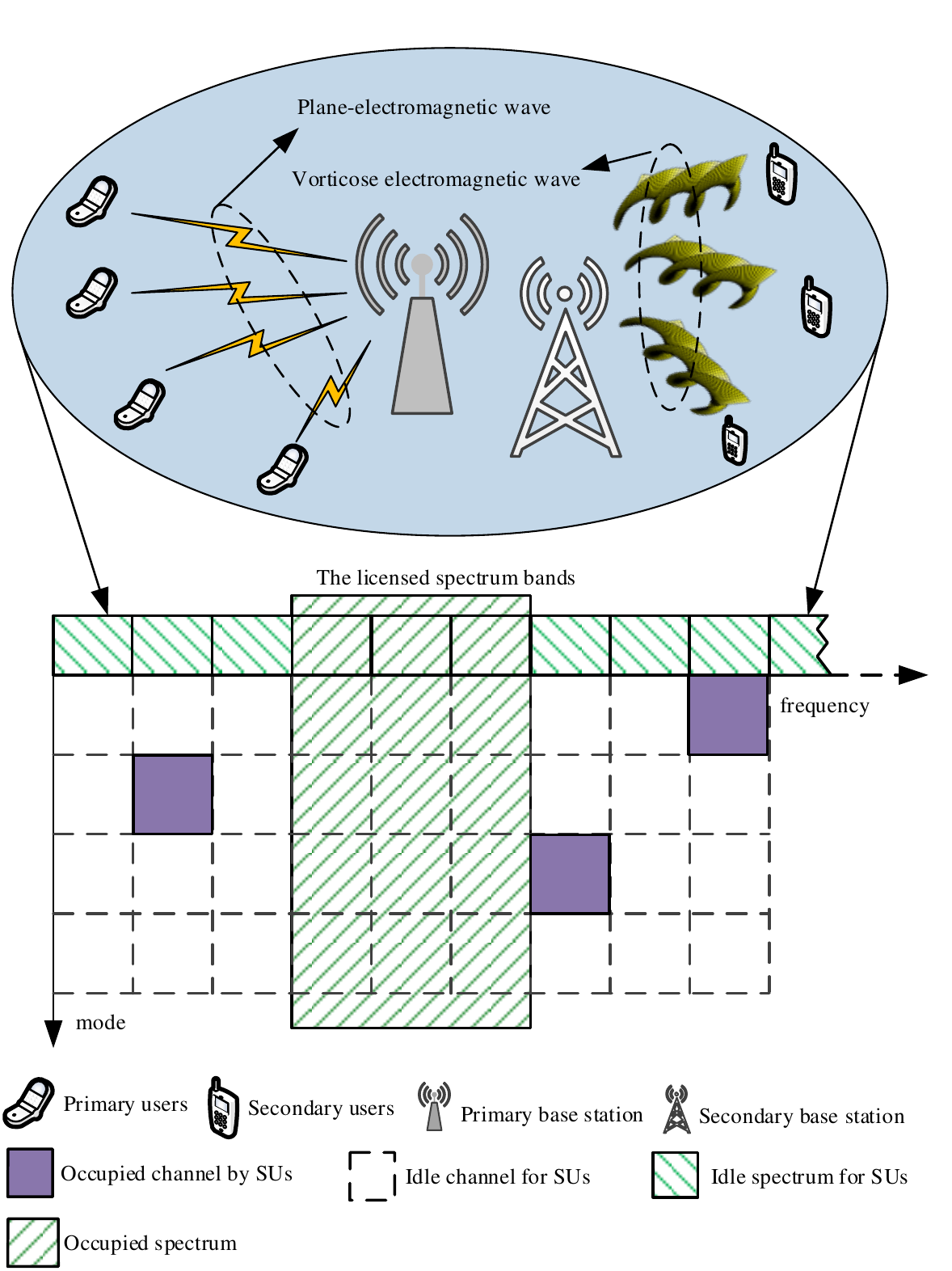}\\
\caption{The system model for cognitive radio networks under mode-frequency hopping.}
\label{fig:sys}%
\vspace{-10pt}
\end{figure}

We consider the cognitive radio network, as depicted in Fig.~\ref{fig:sys}, which consists of the primary base station (BS), the secondary BS, PUs, and SUs. 
The numbers of licensed channels, OAM-modes, PUs, and SUs are denoted by $N$, $L$, $Q$, and $M$, respectively. Theoretically, the number of OAM-modes $L$ can be infinite.
As shown in Fig.~\ref{fig:sys}, PUs transmit signal using the traditional plane-electromagnetic waves in frequency domain while SUs transmit signal using the vorticose electromagnetic waves in both frequency and mode domains. Thus, $NL$ channels can be selected by the SUs to sense and then transmit their own signals within the available channels. All channels are assumed to be independent with each other. PUs have the high priority to access each licensed channel, which follows an ON-OFF model with the channel either busy (ON) or idle (OFF). The probabilities that the states switch from ON to OFF and from OFF to ON are denoted by $\rho$ and $\varrho$, respectively. In addition, PUs and SUs follow the time-slotted structure, where PUs and SUs are synchronized. A SU senses the presence of PUs at the beginning of each time slot and then determines whether to access the licensed channel or not.

We assume that 
the attackers don't know the MFH pattern of SUs and the FH pattern of PUs. Thus, it is difficult for attackers to track the SUs and PUs in each time slot. Also, we assume that the attackers don't cooperate with each other. As a result, a few attackers may jam the same SU at the same time in CRNs. In addition, we assume that the attackers have the same capabilities as SUs to sense and access the licensed channels. The attackers randomly select a channel to jam. SUs cooperate with each other in CRNs. Thus, SUs cannot access or sense the same channel at the same time. 
Due to lack of space, we mainly calculate the false alarm probability, outage probability, and the capacity of SUs.


\section{Anti-Jamming MFH Scheme in CRNs}\label{sec:anti}
In this section, we propose the MFH scheme for anti-jamming in CRNs and derive the corresponding false alarm probability, transmission outage probability, and the capacity of SUs under our proposed MFH scheme. First, we derive the false alarm probability corresponding to the sensing phase and transmission outage probability corresponding to the transmission phase. Then, based on the two derived probabilities, we derive the capacity of SUs for CRNs under our proposed MFH scheme.

\subsection{False Alarm Probability}
We denote by $\mathcal{K}$ the total number of malicious attackers in CRNs and $\varphi$ the azimuthal angle for attackers, PUs, and SUs. Since PUs transmit signals in frequency domain, the order of PUs' transmit signals is considered to be zero. 
Therefore, when the SU senses the idle licensed channel for the 0 OAM-mode corresponding to the $n$-th $(n=0,1,\cdots, N-1)$ band, the SU's sensing signal, denoted by $y_{J}(t)$, can be expressed as follows:
\begin{equation}
    y_{J}(t)=\sum_{k=0}^{\mathcal{K}} h_{k} s_{k}(t) e^{j \varphi l_{k}} e^{j 2\pi f_{k} t}+h_{p}s_{p}(t) e^{j 2\pi f_{n} t}+n(t),
\end{equation}
where $t$ is the time variable, $h_{k}$, $s_{k}(t)$, $l_{k}$, and $f_{k}$ are the channel amplitude gain from the attacker's transmitter to the SU's receiver, the transmit signal at the sensing phase, the OAM-mode, and the carrier frequency, respectively, of the $k$-th $(k=0,1,\cdots, \mathcal{K})$ attacker. $h_{p}$ denotes the channel amplitude gain from the PU's transmitter to the SU's receiver, $s_{p}(t)$ denotes the transmit signal of the PU, $f_{n}$ represents the carrier frequency of the $n$-th band, and $n(t)$ the additive white Gaussian noise (AWGN) with zero mean. When $k=0$, it means that there are no attackers at the sensing phase. Also, we assume $h_{0}=0$ and $s_{0}(t)=0$.
\setcounter{TempEqCnt}{\value{equation}} 
\setcounter{equation}{7} 
\begin{figure*}[ht]
\begin{eqnarray}
   p_{s}(\epsilon,K_{s}, m)\hspace{-0.3cm}&=& \hspace{-0.3cm}{\rm Pr}\left(\sum_{k=1}^{K_{s}} h_{k}^{2} P_{J} + h_{p}^{2} P_{p} \geq \epsilon \sigma^{2}\right)
   \nonumber\\
   \hspace{-0.3cm}&=& \hspace{-0.3cm} 1- \int_{0}^{\frac{\epsilon \sigma^{2}}{P_{J}}}  p_{h}(h) d h  \int_{0}^{\frac{\epsilon \sigma^{2} -P_{J} h }{P_{p}}} p_{h_{p}^{2}}(h_{p}^{2}) d (h_{p}^{2})
   \nonumber\\
   \hspace{-0.3cm}&=& \hspace{-0.3cm} 1- \int_{0}^{\frac{\epsilon \sigma^{2}}{P_{J}}}\frac{m^{mK_{s}} h^{mK_{s}-1}e^{-\frac{mh}{\alpha}}}{\alpha^{m K_{s}} \Gamma(m K_{s})} \left[1-e^{-\frac{m}{\alpha}\left(\frac{\epsilon \sigma^{2}- P_{J} h}{P_{p}}\right)}\right]
   \left\{\sum_{u=0}^{m-1}\frac{1}{\Gamma{(m-u)}}\left[\frac{m}{\alpha}\left(\frac{\epsilon \sigma^{2}- P_{J} h}{P_{p}}\right)\right]^{m-1-u}\right\} d h
   \nonumber\\
    \hspace{-0.3cm}&=& \hspace{-0.3cm} 1-\frac{G(m K_{s},\frac{m \epsilon \sigma^{2}}{\alpha P_{J}})}{ \Gamma(m K_{s})}+ \frac{1}{\Gamma(mK_{s})} e^{-\frac{m\epsilon \sigma^{2}}{\alpha P_{p}}} \sum_{u=0}^{m-1}\sum_{v=0}^{m-1-u} \frac{\left(-\frac{P_{J}}{P_{p}}\right)^{m-u-v-1}}{\Gamma(v-1)\Gamma(m-u-v)} \left(\frac{m \epsilon \sigma^{2}}{\alpha P_{p}}\right)^{v} \nonumber\\
    &&\hspace{3.2cm} \times G\left[mK_{s}+m-u-v-1, \frac{\epsilon \sigma^{2}}{P_{J}\left(\frac{m}{\alpha}-\frac{m P_{J}}{\alpha P_{p}}\right)}\right]  \left(1-\frac{ P_{J}}{P_{p}}\right)^{-mK_{d}-m+u+v+1}.
    \label{eq:integ}
\end{eqnarray}
\hrulefill
\end{figure*}
\setcounter{equation}{\value{TempEqCnt}}

To decompose OAM-mode, the SU's sensing signal $y_{J}(t)$ is multiplied with $e^{-j \varphi l} (l=0,1,\cdots, L-1)$ and then passes through an integrator~\cite{2018_MFH}. We assume that the transmit signals of $\widetilde{K}_{s} (\widetilde{K}_{s}=0,1,\cdots, \mathcal{K})$ attacker carry the same OAM-mode $l$ as that of the SU. Thus, we have
\begin{eqnarray}
    \frac{1}{2\pi} \int_{0}^{2\pi} y_{J}(t) e^{-j \varphi l} d \varphi\hspace{-0.3cm} &=& \hspace{-0.3cm} \!\!\sum_{k=0}^{\widetilde{K}_{s}} h_{k} s_{k}(t)e^{j 2\pi f_{k} t}\!+\!h_{p}s_{p}(t)e^{j 2\pi f_{n} t}\nonumber\\
   \hspace{-0.3cm} && \hspace{-0.3cm} +\frac{1}{2\pi} \int_{0}^{2\pi} n(t)e^{-j \varphi l} d \varphi.
\end{eqnarray}
Then, the signal passes through the band pass filter. Clearly, only when both the carrier frequencies and OAM-modes of attackers' signals are the same with that of the SU required at the sensing phase, the integrator and band pass filter cannot fully cancel the interference.
Otherwise, the interference can be entirely removed by the integrator or band pass filter. We assume that $K_{s} (K_{s}=0,1,\cdots, \mathcal{K})$ attackers jam the SU's transmit signal at each sensing phase after passing through the integrator and band pass filter. Thus, the SU's sensing signal, denoted by $y_{s}(t)$, after passing through the integrator and band pass filter in CRNs under our proposed MFH scheme can be expressed as follows:
\begin{eqnarray}
y_{s}(t)=\sum_{k=0}^{K_{s}}h_{k} s_{k}(t)+h_{p} s_{p}(t)+\tilde{n}(t), \label{eq:decom}
\end{eqnarray}
where $\tilde{n}(t)$ is the noise with zero mean and variance of $\sigma^{2}$ after passing through the band pass filter.

We denote by $P_{J}$ and $P_{p}$ the transmit power of each attacker and the PU, respectively. Based on Eq.~\eqref{eq:decom}, we can calculate the instantaneous SINR of sensing, denoted by $\gamma_{s}$, with $K_{s}$ malicious attackers at the sensing phase in CRNs as follows:
\begin{eqnarray}
    \gamma_{s}=\frac{P_{J}\sum\limits_{k=0}^{K_{s}}h_{k}^{2}+h_{p}^{2}P_{p}}{\sigma^{2}}.\label{eq:INR}
\end{eqnarray}
Generally, SUs are extremely sensitive to the existence of PUs in CRNs, which means that even though a PU occupying the licensed channel has a very low instantaneous SINR, SUs cannot be allowed to access the licensed channel to transmit their own signal. Therefore, the sensing threshold of SU's instantaneous SINR, denoted by $\epsilon$, is highly small in CRNs. If the value of instantaneous SINR is larger than $\epsilon$, it implies that the licensed channel is busy and occupied by a PU or attackers which disguise PUs. Thus, the SU must sense another licensed channel for transmission. If the value of instantaneous SINR is smaller than $\epsilon$, it implies that the SU can access the idle licensed channel.

Nakagami-$m$ fading can be utilized in radio vortex wireless communications, where $m$ is the fading parameter. 
Based on the probability density function (PDF) of channel amplitude gain $h_{k}$, the PDF of $h_{k}^{2}$, denoted by $p_{h_{k}^{2}}(h_{k}^{2})$, can be derived as follows:
\begin{eqnarray}
    p_{h_{k}^{2}}(h_{k}^{2})=\frac{1}{\Gamma(m)}\left(\frac{m}{\alpha}\right)^{m} h_{k}^{2(m-1)} e^{-\frac{m h_{k}^{2}}{\alpha}},
    \label{eq:PDF_HK}
\end{eqnarray}
where $\alpha$ is the expectation with respect to $h_{k}^{2}$ and $\Gamma(\cdot)$ is the Gamma function. 
We denote by $h$ the sum of $h_{k}^{2}$ from 1 to $K_{s}$, i.e. $h=\sum_{k=0}^{K_{s}}h_{k}^{2}$. Since the fadings are mutually statistically independent, $h_{k}^{2}$ are statistically independent. 
Hence, based on the analyses in MFH communications~\cite{2018_MFH}, we can derive the joint PDF of $\sum_{k=0}^{K_{s}}h_{k}^{2}$, denoted by $p_{h}(h)$, as follows:
\begin{eqnarray}
    p_{h}(h)=\frac{m^{mK_{s}} h^{mK_{s}-1}e^{-\frac{mh}{\alpha}}}{\alpha^{m K_{s}} \Gamma(m K_{s})},\ \ h \geq 0.\label{eq:PDF_sen}
\end{eqnarray}
The PDF of $h_{p}^{2}$, denoted by $p_{h_{p}^{2}}(h_{p}^{2})$, can be obtained by replacing $h_{k}^{2}$ by $h_{p}^{2}$ in Eq.~\eqref{eq:PDF_HK}.
Combining Eq.~\eqref{eq:INR} and the condition $\gamma_{s} \geq \epsilon$, we have
\begin{eqnarray}
   \sum_{k=1}^{K_{s}} h_{k}^{2} P_{J} + h_{p}^{2} P_{p} \geq \epsilon \sigma^{2}. \label{eq:epsilon}
\end{eqnarray}
Based on Eqs.~\eqref{eq:PDF_sen} and \eqref{eq:epsilon}, we can derive the false alarm probability, denoted by $p_{s}(\epsilon, K_{s}, m)$, that $K_{s}$ attackers jam the SU's sensing signal for CRNs, as shown in Eq.~\eqref{eq:integ},
where $G(a,b)$ represents the incomplete Gamma function for corresponding elements of $a$ and $b$ and is given by
\setcounter{equation}{8}
\begin{eqnarray}
    G(a,b)=\int_{0}^{b} z^{a-1} e^{-z} d z.
\end{eqnarray}
Assuming that $NL$ licensed channels are idle for the SU, malicious attackers randomly access these idle channels with equal probability. Thus, the probability that an attacker accesses the same licensed channel with the SU is $\frac{1}{NL}$ at the transmission period.
\setcounter{TempEqCnt}{\value{equation}} 
\setcounter{equation}{17} 
\begin{figure*}[ht]
\begin{eqnarray}
   p_{d}(\eta,K_{d},m)\hspace{-0.3cm}&=& \hspace{-0.3cm}\int_{0}^{\infty}  p_{H}(H) d H  \int_{0}^{\frac{\eta P_{J} H+\eta \sigma^{2}}{P_{c}}} p_{h_{s}^{2}}(h_{s}^{2}) d (h_{s}^{2})
   \nonumber\\
   \hspace{-0.3cm}&=& \hspace{-0.3cm} \int_{0}^{\infty}\frac{m^{mK_{d}} H^{mK_{d}-1}e^{-\frac{mH}{\alpha}}}{\alpha^{m K_{d}} \Gamma(m K_{d})} \left[1-e^{-\frac{m}{\alpha}\left(\frac{\eta P_{J} H+\eta \sigma^{2}}{P_{c}}\right)}\right]
   \left\{\sum_{u=0}^{m-1}\frac{1}{\Gamma{(m-u)}}\left[\frac{m}{\alpha}\left(\frac{\eta P_{J} H+\eta \sigma^{2}}{P_{c}}\right)\right]^{m-1-u}\right\}
   d H
   \nonumber\\
  \hspace{-0.3cm} &=& \hspace{-0.3cm} 1 - \frac{1}{\Gamma(mK_{d})} e^{-\frac{m\eta\sigma^{2}}{\alpha P_{c}}} \sum_{u=0}^{m-1}\sum_{v=0}^{m-1-u} \frac{\Gamma(mK_{d}+m-u-v-1)}{\Gamma(v-1)\Gamma(m-u-v)} \left(\frac{\eta P_{J}}{P_{c}}\right)^{m-u-v-1} \nonumber\\
   &&\hspace{8cm} \left(\frac{m\eta \sigma^{2}}{\alpha P_{c}}\right)^{v} \left(1+\frac{\eta P_{J}}{P_{c}}\right)^{-mK_{d}-m+u+v+1}.
   \label{eq:p_eta}
\end{eqnarray}
\hrulefill
\end{figure*}
\setcounter{equation}{\value{TempEqCnt}}

The probability that there are $K_{s}$ malicious attackers at the sensing phase, denoted by $P_{K_{s}}$, in CRNs under our proposed MFH scheme can be calculated as follows:
\begin{eqnarray}
    P_{K_{s}}= \dbinom{\mathcal{K}}{K_{s}}\left(\frac{1}{NL}\right)^{K_{s}}\left(1-\frac{1}{NL}\right)^{\mathcal{K}-K_{s}}.
\end{eqnarray}
When SUs hop to the non-zero OAM-mode, the sensed signal of SU, denoted by $\tilde{y}_{s}(t)$, after passing through the integrator and band pass filter can be obtained as follows:
\begin{eqnarray}
   \tilde{y}_{s}(t)=\sum_{k=0}^{K_{s}}h_{k} s_{k}(t)+\tilde{n}(t).
\end{eqnarray}
Similar to the analyse above, the corresponding false alarm probability that $K_{s}$ attackers jam the SU's sensing signal, denoted by $\tilde{p}_{s}(\epsilon, K_{s}, m)$, in CRNs can be derived as follows:
\begin{eqnarray}
  \tilde{p}_{s}(\epsilon, K_{s}, m)=1-\frac{G(m K_{s},\frac{m \epsilon \sigma^{2}}{\alpha P_{J}})}{ \Gamma(m K_{s})}.
\end{eqnarray}

Therefore, the average false alarm probability, denoted by $P_{s}$, at the sensing phase for all possible $K_{s}$ malicious attackers in CRNs under our proposed MFH scheme can be derived as follows:
\begin{eqnarray}
    P_{s}=\sum_{K_{s}=0}^{\mathcal{K}}P_{K_{s}}\left[\frac{1}{L}p_{s}(\epsilon,K_{s}, m)+\frac{L-1}{L}\tilde{p}_{s}(\epsilon,K_{s}, m)\right].
\end{eqnarray}


\subsection{Transmission Outage Probability}
When the SU accesses the licensed channel, the SU's received signal, denoted by $r_{J}(t)$, for the $l$-th OAM-mode corresponding to the $n$-th band at the receiver can be expressed as follows:
\begin{equation}
    r_{J}(t)=h_{s}x(t)e^{j \varphi l} e^{j 2\pi f_{n} t}+\sum_{k=0}^{\mathcal{K}} h_{k} x_{k}(t) e^{j \varphi l_{k}} e^{j 2\pi f_{k} t}+n(t),
\end{equation}
where $h_{s}$, $x(t)$, and $x_{k}(t)$ denote the channel amplitude gain from the SU's transmitter to the SU's receiver, the transmit signal of the SU, and the transmit signal of the $k$-th attacker at the transmission phase. When $k=0$, we assume that $x_{0}(t)=0$.

Then, similar to the analyses at the sensing phase, the interfering signals with different OAM-modes or carrier frequencies can be filtered after the received signal $r_{J}(t)$ passing through the integrator and band pass filter in CRNs under the MFH scheme. We assume that the SU's transmit signal can be jammed by $K_{d}$ $(K_{d}=0,1,\cdots, \mathcal{K})$ attackers in CRNs under our proposed MFH scheme. Thus, we can obtain the SU's received signal, denoted by $r_{d}(t)$, as follows:
\begin{eqnarray}
r_{d}(t)=h_{s}x(t)+\sum_{k=0}^{K_{d}}h_{k} x_{k}(t)+\tilde{n}(t).
\end{eqnarray}
Therefore, the received instantaneous SINR, denoted by $\gamma_{d}$, with $K_{d}$ attackers at the transmission phase in CRNs under the MFH scheme can be derived as follows:
\begin{eqnarray}
    \gamma_{d}=\frac{P_{c}h_{s}^{2}}{P_{J}\sum\limits_{k=0}^{K_{d}}h_{k}^{2}+\sigma^{2}}, \label{eq:SINR}
\end{eqnarray}
where $P_{c}$ denotes the transmit power of each SU.

Due to the existence of attackers, the SU's received SINR will severely downgrade. The received SINR decreases as the number of attackers increases. If the SU's received SINR is below the transmission instantaneous SINR threshold value, denoted by $\eta$, the SUs' transmit signal cannot be successfully recovered, thus resulting in communication outage for the SU. If $\gamma_{d} \geq \eta$, it means that communications for the SU is successful.

Based on Eq.~\eqref{eq:SINR} and the condition $\gamma_{d} \leq \eta$, we can obtain
\begin{eqnarray}
    \frac{P_{c}h_{s}^{2}}{\sigma^{2}}-\frac{\eta P_{J} H}{\sigma^{2}} \leq \eta,
\end{eqnarray}
where $H=\sum\limits_{k=0}^{K_{d}}h_{k}^{2}$.

The PDF of $h_{s}^{2}$, denoted by $p_{h_{s}^{2}}(h_{s}^{2})$, can be obtained, where we replace $h_{k}^{2}$ by $h_{s}^{2}$ in Eq.~\eqref{eq:PDF_HK}. The PDF of $H$, denoted by $p_{H}(H)$, at the transmission phase can be obtained by replacing $h$ with $H$ and $K_{s}$ with $K_{d}$ in Eq.~\eqref{eq:PDF_sen}, respectively. Also, $h_{s}^{2}$ and $H$ are mutually independent. Thus, the transmission outage probability, denoted by $p_{d}(\eta, K_{d},m)$, that $K_{d}$ attackers jam the SU's transmit signal can be calculated as Eq.~\eqref{eq:p_eta}. Rayleigh fading is a special case of Nakagami-$m$, where $m=1$. Thus, the transmission outage probability $p_{d}(\eta,K_{d},m)$ for Rayleigh fading can be rewritten as follows:
\setcounter{equation}{18}
\begin{eqnarray}
    p_{d}(\eta,K_{d},1)=1-e^{-\frac{\eta \sigma^{2}}{\alpha P_{c}}} \left(1+\frac{\eta \alpha P_{J}}{P_{c}}\right)^{-K_{d}}.
\end{eqnarray}
Also, we can obtain the probability, denoted by $P_{K_{d}}$, that $K_{d}$ malicious attackers jam the SU's transmit signal simultaneously at the transmission phase in CRNs under the MFH scheme as follows:
\begin{eqnarray}
    P_{K_{d}}= \dbinom{\mathcal{K}}{K_{d}}\left(\frac{1}{NL}\right)^{K_{d}}\left(1-\frac{1}{NL}\right)^{\mathcal{K}-K_{d}}.
\end{eqnarray}
When no attackers jam the SU's transmit signal, the transmission outage probability is with the smallest value. We can write the minimum transmission outage probability $p_{d}(\eta,K_{d},m)$ with no attackers in CRNs as follows:
\begin{eqnarray}
    p_{d}(\eta,0,m)=\frac{G\left(m,\frac{m\eta \sigma^{2}}{\alpha P_{c}}\right)}{\Gamma(m)}.
\end{eqnarray}

\subsection{Capacity in CRNs Under Our Proposed MFH Scheme}

To clearly show that our developed MFH scheme in CRNs can achieve better anti-jamming results, we consider to analyze the capacity of SUs. If the licensed channel is detected as occupied by attackers, which disguise PUs, the SU will wait until the licensed channel becomes idle and then transmits signals. 
In the following, we derive the capacity of SUs under our proposed MFH scheme for anti-jamming in CRNs.

The successful transmission probability of a SU, denoted by $p_{suc}(K_{d},m)$, with $K_{d}$ malicious attackers at the transmission phase can be obtained as follows:
\begin{eqnarray}
   p_{suc}(K_{d},m)=(1-P_{s})[1-P_{d}(\eta,K_{d},m)]P_{K_{d}}.
\end{eqnarray}
Then, the instantaneous capacity, denoted by $C_{MFH}$, of the SU for all possible $K_{d}$ can be derived as follows:
\begin{eqnarray}
    C_{MFH}=\sum_{K_{d}=0}^{\mathcal{K}}p_{suc}(K_{d},m)B\log_{2}(1+\gamma_{d}),
\end{eqnarray}
where $B$ represents the bandwidth of each channel. For Nakagami-$m$ fading, the PDF of SINR, denoted by $p_{\gamma}(\gamma)$, can be expressed as follows:
\begin{eqnarray}
   p_{\gamma}(\gamma)=\frac{1}{\Gamma(m)}\left(\frac{m}{\bar{\gamma}}\right)^{m} \gamma^{m-1} e^{-\frac{m \gamma}{\bar{\gamma}}},
\end{eqnarray}
where $\gamma$ denotes the SINR of channel and $\bar{\gamma}$ denotes the average SINR.

Therefore, we can obtain the capacity, denoted by $C$, of all SUs for all possible $K_{d}$ malicious attackers in CRNs under our proposed MFH scheme as follows:
\begin{eqnarray}
    C= M B \sum_{K_{d}=0}^{\mathcal{K}}p_{suc}(K_{d}) \mathbb{E}_{\gamma}\left[\log_{2}(1+\gamma_{d})\right],
\end{eqnarray}
where $\mathbb{E}_{\gamma_{d}}(\cdot)$ represents the expectation operation with respect to $\gamma_{d}$.

\section{Performance Evaluation}\label{sec:perfor}
In this section, we evaluate the performances of our proposed scheme and compare the capacities of our developed MFH scheme with the conventional FH scheme in CRNs. Throughout our evaluations, we set the number of available carriers as 2, the transmit power of each attackers as 0.1 W, the number of SUs as 4, and the bandwidth as 10 MHz.

%
%

\begin{figure}
\centering
\includegraphics[scale=0.65]{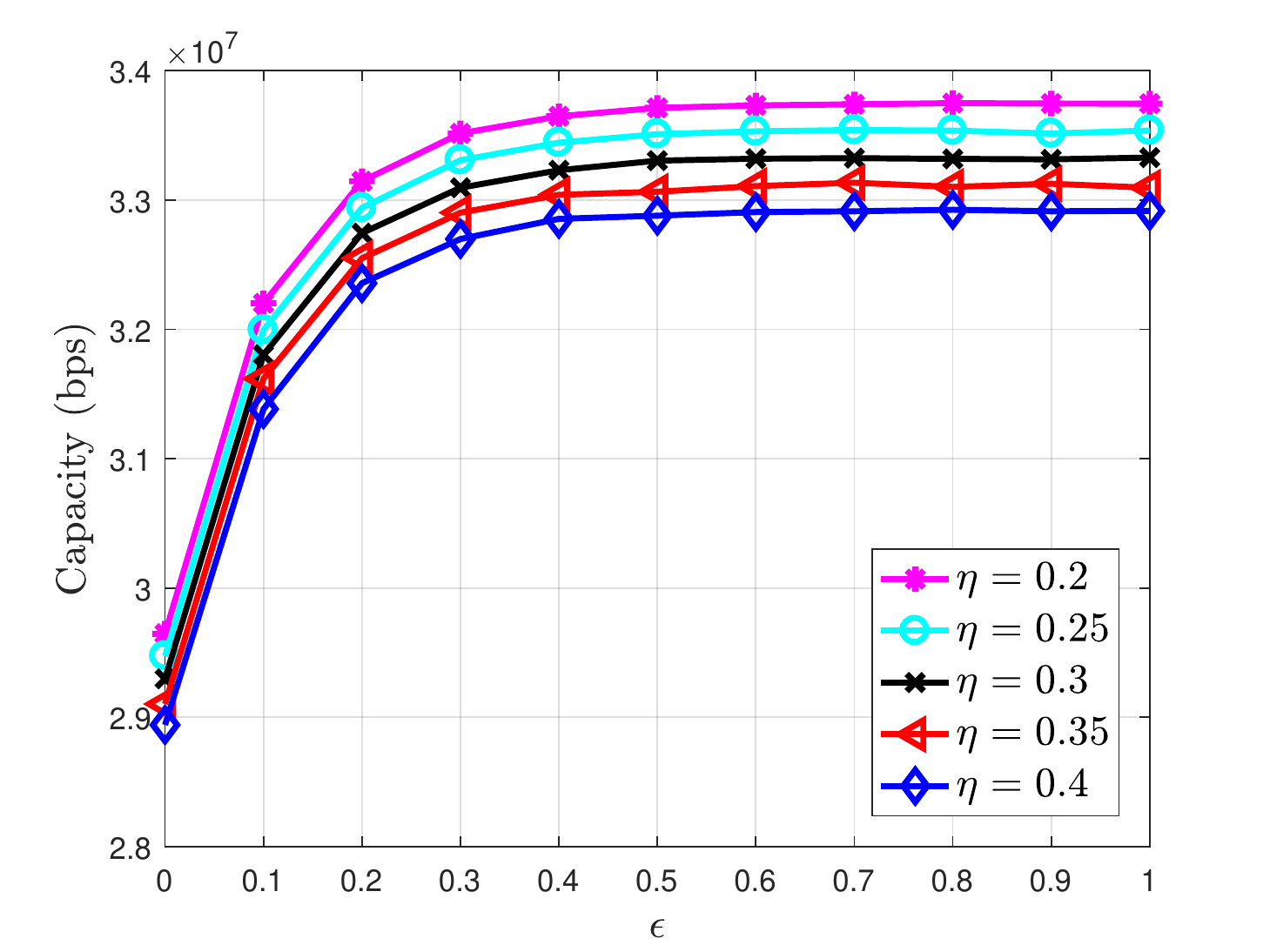}\\
\caption{The capacity of SUs versus the SINR sensing threshold $\epsilon$ in CRNs under our proposed MFH scheme.}
\label{fig:capacity_epsilon}
\end{figure}

Figure~\ref{fig:capacity_epsilon} shows the capacity of SUs versus the instantaneous SINR sensing threshold $\epsilon$ in CRNs under our proposed MFH scheme, where we set the total number of attackers as 2, the number of available OAM-modes as 8, and the transmission threshold $\eta$ as 0.2, 0.25, 0.3, 0.35, and 0.35, respectively. As shown in Fig.~\ref{fig:capacity_epsilon}, the capacity of SUs increases as the  instantaneous SINR sensing threshold increases and transgression threshold decreases. This is because the false alarm probability at the sensing phase decreases as $\epsilon$ increases, thus resulting in the increase of the probability that SUs access the licensed channel. Also, the transmission outage probability at the transmission phase decreases as $\eta$ decreases. Therefore, the successful transmission probability of SUs increases. In addition, we can see that the capacity of SUs is very close to a fixed value as $\epsilon$ increases. 
These results prove that we can get large SUs' capacity with small  transmission threshold and big instantaneous SINR sensing threshold in CRNs under our developed MFH scheme.
\begin{figure}
\centering
\includegraphics[scale=0.65]{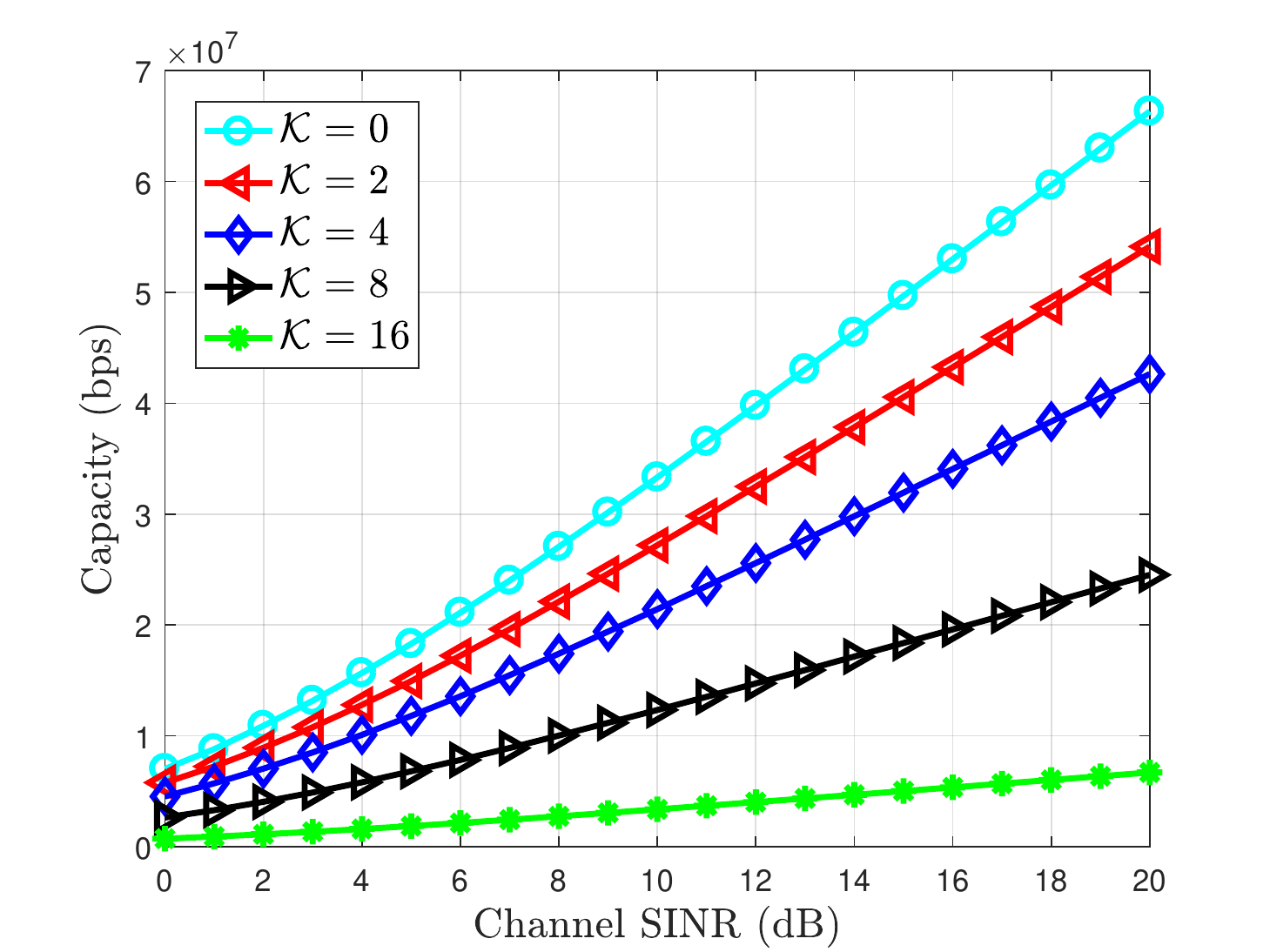}\\
\caption{The capacity of SUs versus the channel SINR with respect to different number of malicious attackers in CRNs under the MFH scheme.}
\label{fig:capacity_K}
\vspace{-10pt}
\end{figure}

\begin{figure}
\centering
\includegraphics[scale=0.65]{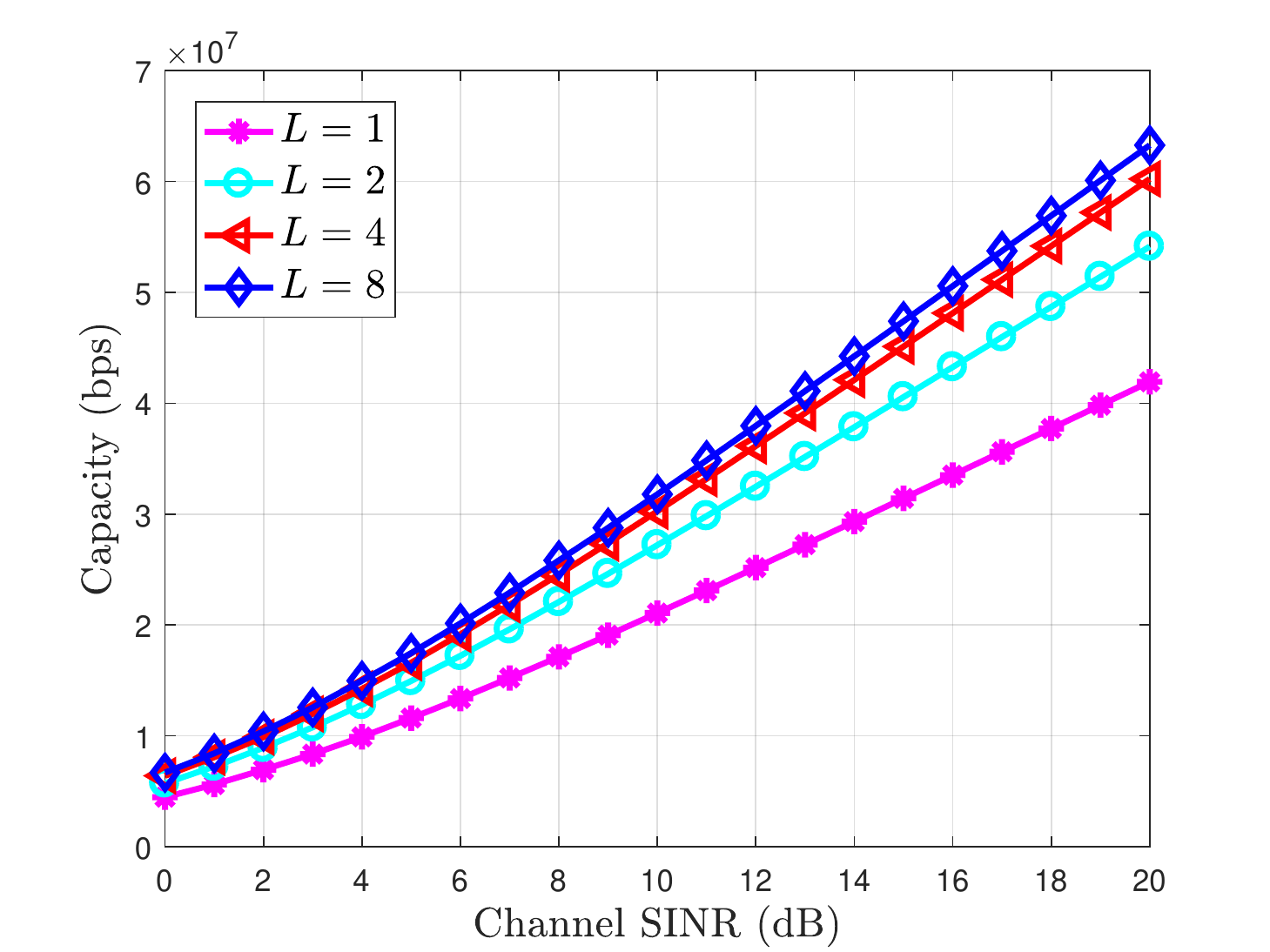}\\
\caption{The capacity comparison between our developed MFH scheme and the conventional FH in CRNs versus the channel SINR.}
\label{fig:capacity_L}
\vspace{-10pt}
\end{figure}

Figure~\ref{fig:capacity_K} depicts the capacity of SUs versus the channel SINR with respect to different number of malicious attackers in CRNs under the MFH, where we set the instantaneous SINR sensing  threshold $\epsilon$ as 0.1, transmission threshold $\eta$ as 0.3, the number of available OAM-modes as 2, and the number of attackers as 0, 2, 4, 8, and 16, respectively. When the number of attackers is zero, it means that SU's sensing and transmission work without interference from attackers. The capacity of SUs increases as the channel SINR increases. Also, the capacity of SUs decreases as the number of attackers increases. The increase of attackers leads to the increase of SINR at the sensing phase and decrease of SINR at the transmission phase. Hence, the false alarm probability at the sensing phase and the transmission outage probability at the transmission phase increase, which also results in the decrease of capacity for SUs. Therefore, CRNs under our developed MFH scheme can achieve high capacity with the decrease of attackers in high channel SINR region.

Figure~\ref{fig:capacity_L} compares the capacities of SUs between our developed MFH scheme and the conventional FH schemes in CRNs, where we set the instantaneous SINR sensing threshold $\epsilon$ as 0.1, transmission threshold $\eta$ as 0.3, the number of attackers as 2, and the number of available OAM-modes as 1, 2, 4, and 8, respectively. Observing Fig.~\ref{fig:capacity_L}, we can see that the capacity increases as the number of OAM-modes increases. The reason is that the increase of OAM-modes makes malicious attackers  difficult to track the SUs, thus reducing the probability to jam the SUs. When there is only one OAM-mode to hop and the order of the OAM-mode is zero, it means that our developed MFH scheme is equivalent to the conventional FH scheme in CRNs. Clearly, the capacity of SUs using our developed MFH scheme is larger than that of using the conventional FH in CRNs. These results prove that our developed MFH scheme can achieve better anti-jamming results than the conventional FH in CRNs.

\section{Conclusions} \label{sec:conc}

In this paper, we proposed to use the MFH scheme for better anti-jamming results in CRNs. We derived the false alarm probability corresponding to the sensing phase and the transmission outage probability corresponding to the transmission phase. Based on the two derived probabilities, we calculated the capacity of SUs in CRNs under the MFH scheme. Numerical results show that the capacity of SUs using the MFH scheme is higher than that of using the conventional FH scheme in CRNs. 

\end{document}